\documentclass[aps,prl, amsmath,amssymb, twocolumn,
 groupedaddress,
 showpacs]{revtex4-1}
\usepackage{bm}
\usepackage{graphicx}
\usepackage{gensymb}
\usepackage{amsmath}
\usepackage{amssymb}
\usepackage{color}
\usepackage{here}

\begin{document}

\title{Metal-insulator transition and tunable Dirac-cone surface state in the topological insulator TlBi$_{1-x}$Sb$_x$Te$_2$ studied by angle-resolved photoemission}

\author{Chi Xuan Trang,$^1$ Zhiwei Wang,$^{2,3}$ Keiko Yamada,$^1$ Seigo Souma,$^4$ Takafumi Sato,$^1$ Takashi Takahashi,$^{1,4}$ Kouji Segawa,$^5$ and Yoichi Ando$^{2,3}$}

\affiliation{$^1$Department of Physics, Tohoku University, Sendai 980-8578, Japan\\
$^2$Institute of Scientific and Industrial Research, Osaka University, Ibaraki, Osaka 567-0047, Japan\\
$^3$Institute of Physics II, University of Cologne, K$\ddot{o}$ln 50937, Germany\\
$^4$WPI Research Center, Advanced Institute for Materials Research, Tohoku University, Sendai 980-8577, Japan\\
$^5$Department of Physics, Kyoto Sangyo University, Kyoto 603-8555, Japan}

\date{\today}

\begin{abstract}
    We report a systematic angle-resolved photoemission spectroscopy on topological insulator (TI) TlBi$_{1-x}$Sb$_x$Te$_2$ which is bulk insulating at 0.5 $\lesssim x \lesssim$ 0.9 and undergoes a metal-insulator-metal transition with the Sb content $x$. We found that this transition is characterized by a systematic hole doping with increasing $x$, which results in the Fermi-level crossings of the bulk conduction and valence bands at $x$ {$\sim$} 0 and {$x$} $\sim$ 1, respectively. The Dirac point of the topological surface state is gradually isolated from the valence-band edge, accompanied by a sign reversal of Dirac carriers. We also found that the Dirac velocity is the largest among known solid-solution TI systems. The TlBi$_{1-x}$Sb$_x$Te$_2$ system thus provides an excellent platform for Dirac-cone engineering and device applications of TIs.
\end{abstract}

\maketitle

 The three-dimensional topological insulator (3D TI) is a novel quantum state of matter with a metallic surface state and an insulating bulk. The surface state is characterized by linearly dispersive Dirac-cone band dispersion which traverses the bulk-band gap generated by the spin-orbit coupling \cite{AndoReview, SCZhangReview, HasanReview}. Experimental realization of various exotic topological phenomena, such as the topological magnetoelectric effect and anomalous quantum Hall effect, \cite{ME, ME2, AQHE} as well as the device applications of TIs largely rely on the dominance of surface Dirac transport and the capability to manipulate Dirac-carrier properties, whereas it is difficult to achieve highly insulating bulk in most of known the 3D TIs due primarily to the small bulk-band gap and defects in the crystal.

One of the promising strategies to achieve insulating bulk is to reduce defects \cite{RenPRB2010, Cava2012} and compensate defect-induced carriers by tuning chemical composition in a solid-solution system. This methodology has been successfully applied to bulk crystals of tetradymite Bi$_{2-x}$Sb$_x$Te$_{3-y}$Se$_y$ (BSTS) where special combinations of $x$ and $y$ yield high resistivity \cite{TaskinPRL, RenPRB2011, LeePRB2012, JiaPRB2012}. Angle-resolved photoemission spectroscopy (ARPES) study of BSTS has revealed a sign reversal of Dirac carriers while keeping insulating bulk \cite{ArakaneNC}. Discovery of the BSTS system has accelerated the engineering of Dirac carriers and device applications \cite{SulaevNL2015, FatemiPRL2014, OuNC2014, ShiomiPRL2014, AndoNL2014, XuNP2014, KimPRL2014, XiaPRB2013, LeePRB2012}, as highlighted by the spin injection via spin-pumping technique \cite{ShiomiPRL2014}, electrical detection of spin polarization \cite{AndoNL2014}, and realization of quantum Hall effect at high temperature \cite{XuNP2014}. To further realize exotic quantum phenomena and device applications in TIs, it is of particular importance to explore the new materials platform of bulk-insulating TIs, since the essential characteristics of bulk and surface, such as the bulk-band gap and the Dirac velocity, strongly depend on the material.

Rhombohedral III-V-VI$_2$ ternary chalcogenide TlM'X$_2$ [M = Bi and Sb; X = S, Se, and Te; see Figs. 1(a) and 1(b) for the crystal structure and Brillouin zone (BZ), respectively] \cite{YanEPL2010, LinPRL2010, SatoTBSPRL, KurodaTBSPRL, ShenTBTPRL} is an intriguing platform to achieve insulating TIs since its crystal structure is distinct from tetradymite due to the covalent-bonding nature and the absence of van der Waals gap. However, naturally grown crystals are known to be inevitably carrier doped, as seen from the band crossing of bulk conduction band (CB) in TlBiSe$_2$ and TlBiTe$_2$ \cite{SatoTBSPRL, KurodaTBSPRL, ShenTBTPRL}. This problem has hindered an extraction of intrinsic Dirac-carrier properties in TlM'X$_2$. Recently, an attempt has been made to achieve the bulk-insulating phase by fabricating non-stoichiometric TlBiSe$_2$ crystal with excess Bi and deficient Tl \cite{EguchiPRB2014, KurodaPRB2015}. It would, however, be more useful to utilize solid solution in TlM'X$_2$, to take full advantage of this system for realizing novel topological phenomena and device applications.

   In this paper, we report on an ARPES study of TlBi$_{1-x}$Sb$_x$Te$_2$ (TBST) solid solutions. Besides the tunable Dirac carriers and bulk-insulating characteristics, we found that the chemical potential is variable up to $\sim$0.6 eV with varying $x$. The value of $\sim$0.6 eV is the largest among known 3D TIs. In addition, the Dirac velocity is the highest among TI solid solutions. We discuss implications of these findings in comparison with the electronic states of other TIs.
    
 \begin{figure}
 \includegraphics[width=3.4in]{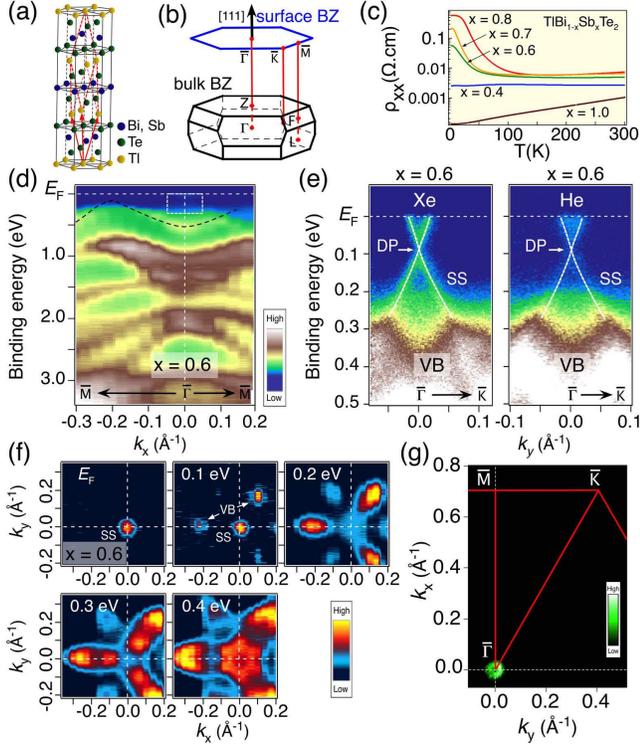}
\caption{(color online) (a) Crystal structure of TBST. (b) Bulk BZ (black lines) and corresponding (111) surface BZ (blue lines). (c) Temperature dependence of electrical resistivity $\rho_{xx}$ for various $x$. (d) Plot of VB ARPES intensity of TBST at $x$ = 0.6 as a function of wave vector ($k_x$) and binding energy ($E_{\rm B}$) measured along the $\bar{\Gamma}\bar{M}$ cut with the He-I$\alpha$ line at 30 K. Black dashed curve is a guide for the eyes to trace the topmost VB. White rectangle corresponds to the $k$ region where the surface state is observed. (e) Near-$E_{\rm F}$ ARPES intensity at $x$ = 0.6 measured along the $\bar{\Gamma}\bar{M}$ cut with the Xe-I and He-I$\alpha$ lines. SS, DP, and VB denote the surface state, the Dirac point, and the valence band, respectively. (f) ARPES-intensity contour plots as a function of 2D wave vector at various  $E_{\rm B}$'s. (g) ARPES intensity mapping at $E_{\rm F}$ covering the entire first BZ.}

\end{figure}
 
   Single-crystalline samples of TBST were grown by a modified Bridgman method, using high-purity elemental shots of Tl (99.99$\%$), Bi (99.9999$\%$), Sb (99.9999$\%$), and Te (99.9999$\%$) as starting materials. To obtain high-quality crystals, we performed surface cleaning to remove the oxide layers formed in air on the raw shots of starting materials \cite{WangAPLMater}. The electrical resistivity shown in Fig. 1(c) exhibits insulating behavior at $x$ = 0.6$-$0.8 in contrast to the metallic behavior at $x$ = 0.4 and 1.0, indicating metal-insulator transitions at $x$ {$\sim$} 0.5 and {$x$} $\sim$ 0.9. It is noted that the plateau in the temperature dependence of the resistivity at low temperature seen for $x$ = 0.7 and 0.8 is likely an indication of a surface conduction channel \cite{RenPRB2010}. ARPES measurements have been performed with a MBS-A1 electron analyzer with He and Xe discharge lamps and a toroidal/spherical grating monochromator. The He-I$\alpha$ ($h\nu$ = 21.218 eV) line and one of the Xe-I ($h\nu$ = 8.437 eV) lines were used to excite photoelectrons. Samples were cleaved $in$-$situ$ along the (111) crystal plane at 5 $\times$ 10$^{-11}$ Torr, and kept at the same temperature during the measurement (see also Sec. 4 of the Supplemental Material \cite{SM} for the cleaving plane). The energy and angular resolutions were set at 2$-$15 meV and 0.2$^\circ$, respectively.   
     
 \begin{figure*}
 \includegraphics[width=6.8in]{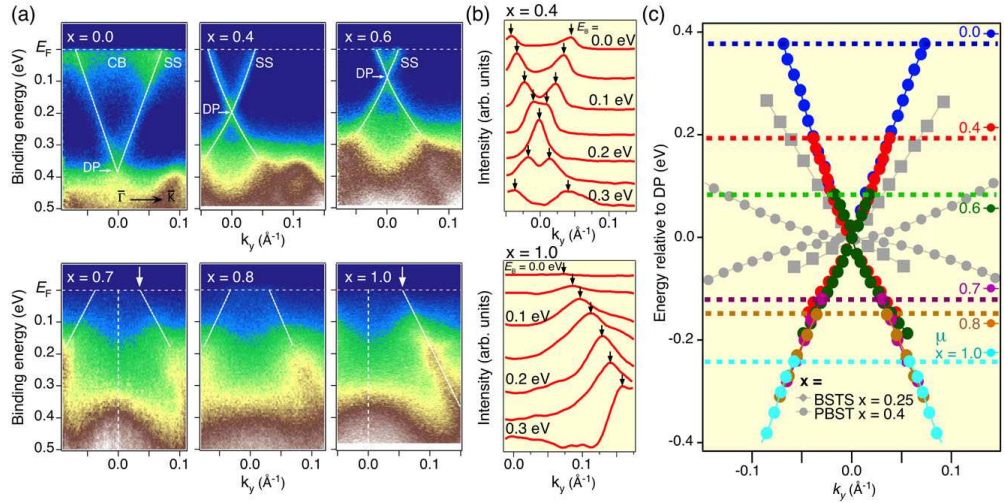}
\caption{(color online). (a) Comparison of near-$E_{\rm F}$ ARPES intensity along the $\bar{\Gamma}\bar{K}$ cut for various $x$ measured with the Xe-I photons at $T$ = 30 K. White dashed curves are a guide for the eyes to trace the Dirac-cone band. CB denotes the bulk conduction band. While arrows for $x$ = 0.7 and 1.0 indicate the location of Fermi wave vector ($k_{\rm F}$). (b) MDCs at selected $E_{\rm B}$'s for (top) $x$ = 0.4 and (bottom) $x$ = 1.0. Arrows indicate the peak position of the MDCs. (c) Composition dependence of the Dirac-cone band dispersion relative to the DP of TBST, compared with those of BSTS for ($x,y$) = (0.25, 1.15) (gray squares) \cite{ArakaneNC} and Pb(Bi$_{1-x}$Sb$_x$)$_2$Te$_4$ (PBST; $x$ = 0.4) (gray circles) \cite{SoumaPBSTPRL}. Energy position of the chemical potential $\mu$ for each composition is indicated by horizontal dashed lines.}

\end{figure*}

First, we present the electronic structure of TBST at $x$ = 0.6 which is in the bulk-insulating phase [see Fig. 1(c)]. We show in Fig. 1(d) the valence-band (VB) ARPES intensity along the $\bar{\Gamma}\bar{M}$ cut in the surface BZ measured with the He-I$\alpha$ line. One can recognize several dispersive bands around $\bar{\Gamma}$, which indicate the high quality of sample surface. The topmost VB shows the ``m''-shaped dispersion and has a top of its dispersion at {$k_x$} $\sim-0.2$ \AA$^{-1}$. One may also see a weak intensity near $E_{\rm F}$ around  $\bar{\Gamma}$. To see this faint feature more clearly, we have performed ARPES measurements with higher accuracy using both the Xe-I and He-I$\alpha$ lines. As is visible in Fig. 1(e), the spectral features are essentially similar between two photon energies; one can see an ``x''-shaped band dispersion arising from the topological Dirac-cone surface state (SS), as confirmed by the $h\nu$ invariance of its energy location. The Dirac point (DP) of the Dirac-cone band is located at the binding energy ($E_{\rm B}$) of $\sim$0.1 eV, indicating the $n$-type nature of Dirac carriers (here, $n$/$p$-type Dirac carrier means Dirac electrons/holes). The observed $E_{\rm {DP}}$ value is much closer to $E_{\rm F}$ compared to those of as-grown thallium-based ternary TIs such as TlBiSe$_2$ (0.3$-$0.4 eV \cite{SatoTBSPRL, KurodaTBSPRL, ShenTBTPRL}) and TlBiTe$_2$ (0.3 eV \cite{ShenTBTPRL}). The reduced electron-doping effect in TBST is likely due to compensation of excess electron carriers by holes created by the (Bi,Sb)/Te antisite defects \cite{RenPRB2011} (see Sec. 1 of Supplemental Material \cite{SM} for the relationship between defects and chemical potential). Besides the proximity of the DP to $E_{\rm F}$, we found no evidence for the appearance of the bulk CB, consistent with the bulk-insulating nature [see Fig. 1(c)]. The absence of the CB in the spectra was also confirmed by the ARPES-intensity mapping covering the entire first BZ in Fig. 1(g), which signifies the single $\bar{\Gamma}$-centered Fermi surface of surface origin. To assess whether the surface transport with the chemical potential located near the DP is possible while keeping an insulating bulk, it is necessary to clarify whether the energy location of the DP is above or below the VB top \cite{ArakaneNC}. For this purpose, we show in Fig. 1(f) the ARPES-intensity map as a function of a two-dimensional (2D) wave vector for various $E_{\rm B}$'s. As is clearly visible, a circular intensity pattern arising from the SS appears at $E_{\rm B}$ = 0.0 ($E_{\rm F}$) and 0.1 eV. Upon increasing $E_{\rm B}$, a signature of the bulk VB first appears at $E_{\rm B}$ = 0.1 eV along the $\bar{\Gamma}\bar{M}$ line as a small circular intensity spot. On further increasing $E_{\rm B}$, the spot gradually expands and evolves into a larger elongated spot, reflecting the holelike nature of the VB.
      
 \begin{figure}
 \includegraphics[width=3.4in]{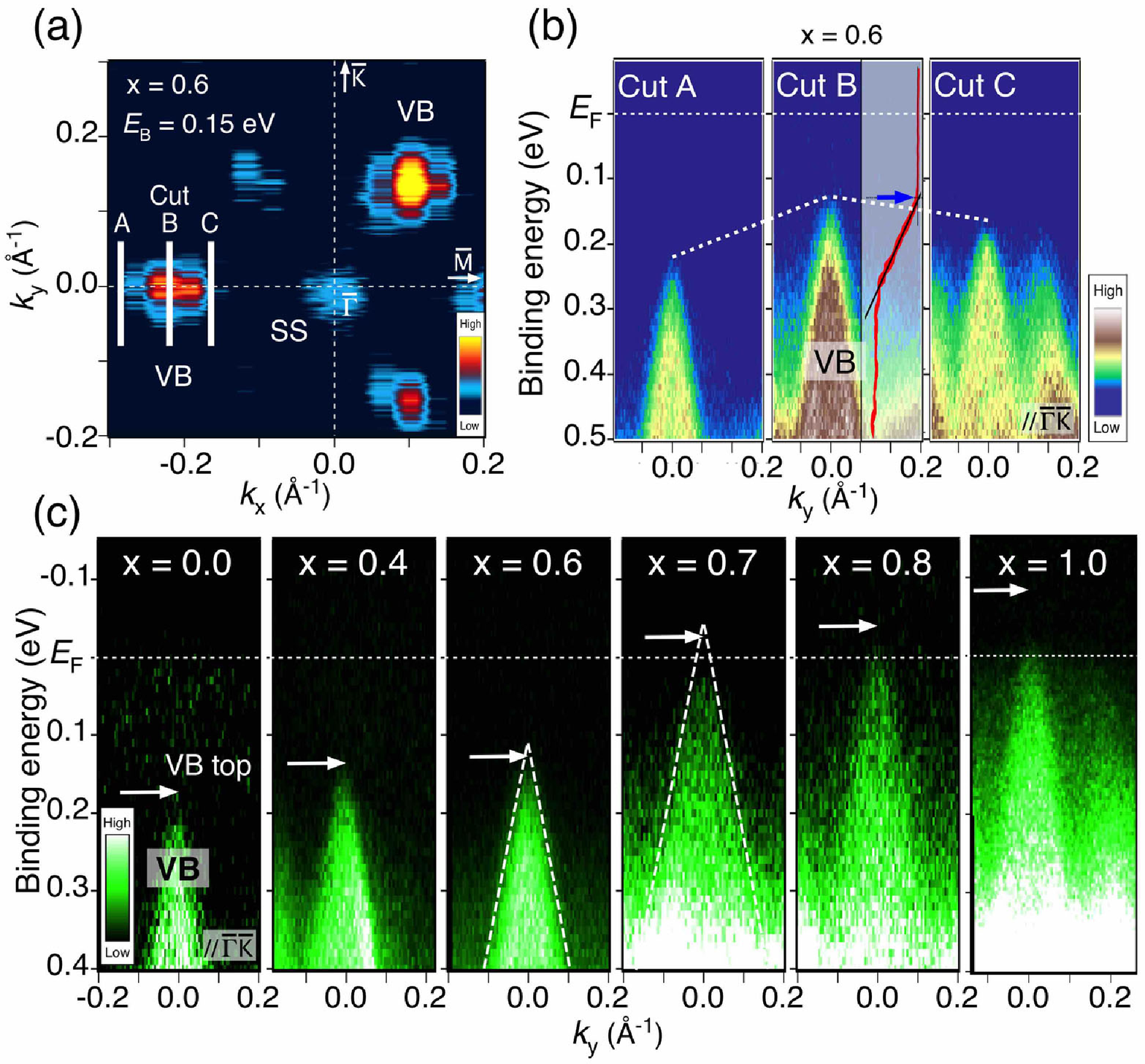}
 \caption{(color online). (a) ARPES-intensity map at $E_{\rm B}$ = 0.15 eV around $\bar{\Gamma}$ for $x$ = 0.6 measured with the He-I photons. (b) Near-$E_{\rm F}$ ARPES intensity measured along cuts A-C shown in panel (a), which highlights the determination of the VB top. White dashed lines connect the local maxima of the VB for each cut. For cut B, the EDC at $k_y$ = 0.0 \AA$^{-1}$ is shown by the red curve. The intersection of two black solid lines on the EDC, which represent the leading edge and the background, corresponds to the location of the VB top (blue arrow) \cite{SatoTBSPRL, ArakaneNC}. (c) Composition dependence of the ARPES intensity for cuts crossing the VB top, highlighting the systematic hole doping into the VB. White dashed lines for $x$ = 0.6 and 0.7 are linear extrapolation of the VB-intensity edges. White arrows indicate the VB top.}
\end{figure}

To see the evolution of the electronic states upon Sb substitution, we have performed systematic ARPES measurements along the $\bar{\Gamma}\bar{K}$ cut for samples with various $x$ values. Results are displayed in Fig. 2(a). On decreasing $x$ from 0.6, one can recognize a gradual downward shift of the Dirac-cone band; the DP for $x$ = 0.4 is located at $E_{\rm B}$ {$\sim$} 0.2 eV, which can be better seen from the momentum distribution curves (MDCs) in Fig. 2(b). As visible in Fig. 2(a), the Dirac cone is further pushed downward at $x$ = 0.0, and it is located at 0.4 eV below $E_{\rm F}$, consistent with the previous study \cite{ShenTBTPRL}. In addition, the bulk CB is observed inside the upper Dirac cone, reflecting the bulk metallic nature. On increasing $x$ from 0.6 to 0.7, the DP is immediately pushed upward into the unoccupied region. This definitely reveals the $n$- to $p$-type transition of Dirac carriers between $x$ = 0.6 and 0.7. For $x\geq0.7$, one can see a systematic hole doping from a comparison of Fermi wave vectors ($k_{\rm F}$'s) for the lower Dirac cone between $x$ = 0.7 and 1.0 (see arrows) [note that the $k_{\rm F}$ points were estimated by tracing the peak position of the MDCs; see the bottom panel of Fig. 2(b)]. These results demonstrate tunable Dirac carriers in TBST, while such tunable nature has been achieved only in a small number of TIs \cite{RenPRB2011, ArakaneNC, SoumaPBSTPRL, ZhangNC2011, YoshimiNC2015}.
 
 \begin{figure}
\includegraphics[width=3.4in]{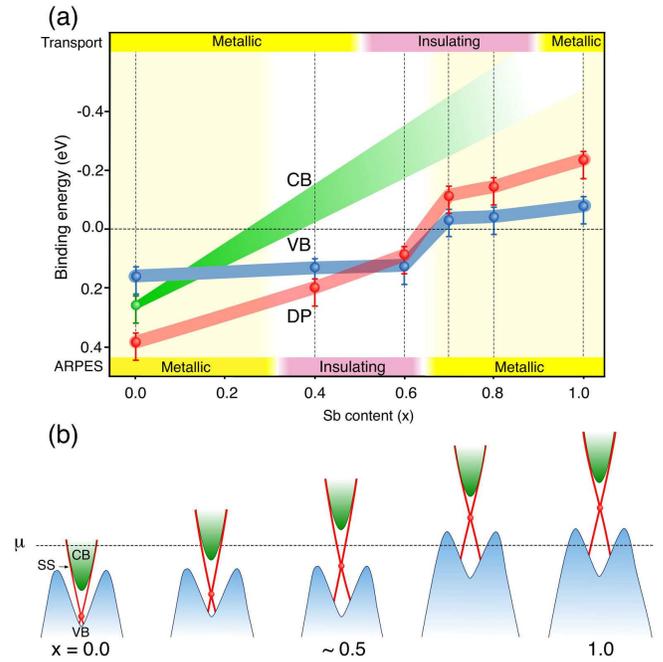}
\caption{(color online). (a) Plot of three characteristic energies, i.e., the conduction-band (CB) bottom (green), the VB top (blue), and the DP energy (red) as a function of $x$. The bulk-insulating/metallic region determined from the resistivity measurements is indicated at the top, while the surface insulating/metallic region suggested from the present ARPES experiment is shown at the bottom. (b) Schematic band diagram of TBST determined from the present ARPES result. Dashed line corresponds to the chemical potential.}
 \end{figure}
 
To quantitatively evaluate the evolution of the SS, we plot in Fig. 2(c) the Dirac-cone dispersion determined from the peak position of the energy distribution curves (EDCs) (see Sec. 2 of Supplemental Material \cite{SM} for the band-dispersion plots). One can see that the energy shift of the Dirac-cone band proceeds in a rigid-band manner; the bands for different $x$ values essentially overlap with each other when we plot its energy position with respect to the DP energy, despite the total chemical-potential ($\mu$) shift of $\sim$0.6 eV. This value is much larger than those of other bulk-insulating TIs such as BSTS ($\sim$0.3 eV) \cite{ArakaneNC} and (Bi$_{1-x}$Sb$_x$)$_2$Te$_3$ thin films ($\sim$0.4 eV) \cite{ZhangNC2011}. As shown in Fig. 2(c), the chemical potential which lies slightly above DP at $x$ = 0.6 is suddenly pushed downward below DP at $x$ = 0.7, again pointing to a sign reversal of Dirac carriers from $n$- to $p$-type at $x$ {$\sim$} 0.65. Such a jump may be explained by the absence of impurity levels within the band gap; in such a case, the impurity level cannot pin the chemical potential, and the chemical potential suddenly jumps from the CB bottom to the VB top with hole doping. We have estimated the Dirac velocity of TBST along the $\bar{\Gamma}\bar{K}$ cut from the slope of the band dispersion around the DP to be 5.4 eV\AA. This value is much larger than those of typical solid-solution TI systems such as BSTS (2.9 eV\AA) \cite{ArakaneNC} and Pb(Bi,Sb)$_2$Te$_4$ (PBST; 0.5 eV\AA) \cite{SoumaPBSTPRL}, as is visible in Fig. 2(c).

 The next important issue is to pin down the location of the bulk-band edges, which is crucial for the Dirac transport properties. As seen from Fig. 2(a), the CB which is seen at $x$ = 0.0 looks absent at $x$ = 0.4$-$1.0 due to hole doping. It is thus suggested that the CB is occupied only in the $x$ region around 0.0 (see Sec. 3 of Supplemental Material \cite{SM} for the estimation of the CB bottom). We used the He-I$\alpha$ line to estimate the VB top, since the corresponding $k_z$ value is close to the $Z$ point of the bulk BZ around which the VB top is located \cite{ShenTBTPRL, KurodaTBSPRL, HasanTBSScience}. As seen from the ARPES-intensity map at $x$ = 0.6 in Fig. 3(a) measured with the He-I$\alpha$ line, the VB top is seen as a circular intensity spot away from $\bar{\Gamma}$ along the $\bar{\Gamma}\bar{M}$ line [see also Fig. 1(f)]. This demonstrates the indirect nature of the bulk-band gap in TBST like in many other TIs \cite{SatoTBSPRL, KurodaTBSPRL, ShenTBTPRL, ArakaneNC, AndoReview, HasanReview}. To accurately determine the VB top, we have measured the ARPES data at several cuts, and show the representative results in Fig. 3(b). As shown in Fig. 3(b), the holelike VB which has a top of dispersion at $E_{\rm B} \sim$ 0.2 eV in cut A moves upward in cut B and then moves downward in cut C, indicating that the VB top is located along cut B at $k_y$ = 0.0 \AA$^{-1}$. We have estimated the VB top to be $E_{\rm B} \sim$ 0.12 eV from the leading edge of the EDC \cite{SatoTBSPRL, ArakaneNC}. This value is slightly higher than the DP energy of $\sim$0.08 eV [see Fig. 2(c)], suggesting that the Dirac cone is isolated from the bulk VB at $x$ = 0.6. This isolation is a prerequisite for realizing some of the exotic topological phenomena and device applications \cite{AndoReview, SCZhangReview, HasanReview}. As displayed in Fig. 3(c), we have systematically traced the VB top for all range of $x$. Obviously, the VB top is located at well below $E_{\rm F}$ for $x$ = 0.0$-$0.6 with a small upward shift upon increasing $x$. The VB top appears to show a sudden upward shift toward the above-$E_{\rm F}$ region at $x$ = 0.7 [note that the VB top for $x$ = 0.7$-$1.0 can be approximately estimated from the crossing point of the linearly extrapolated intensity edges (e.g. see dashed lines for $x$ = 0.6 and 0.7) since the VB top estimated from the EDC almost coincides with this crossing point]. On further increasing $x$, the VB top moves further upward and is finally located at $\sim$0.1 eV above $E_{\rm F}$ at $x$ = 1.0.
 
From these ARPES results, we now discuss the location of the bulk bands with respect to the chemical potential. We summarize in Fig. 4(a) the $x$ dependence of characteristic energy scales, DP energy, VB top, and CB bottom. At $x$ = 0.0, the CB bottom is situated at an energy lower than the VB top, signifying the semimetallic bulk-band structure with a negative band gap, as illustrated in Fig. 4(b), consistent with the previous ARPES study of TlBiTe$_2$ \cite{ShenTBTPRL}. At $x$ = 0.4, the relative position of the VB top and the CB bottom is reversed. Moreover, the chemical potential lies within the band gap. It is noted that unlike BSTS \cite{ArakaneNC}, it turned out to be difficult to determine the CB bottom above $E_{\rm F}$ for $x\geq0.4$ since the surface chemical potential was robust against intentional surface aging (see Sec 4 of the Supplemental Material \cite{SM} for the surface-aging experiments and the cleaving plane \cite{KurodaPRB, NC}). We thus draw a plausible location of the CB bottom with a reasonable ambiguity. As shown in Fig. 4(a), the bulk-insulating character persists up to $x$ {$\sim$} 0.65 at which the VB starts to cross $E_{\rm F}$. Therefore, from the spectroscopic point of view, one can say that (i) the metallic bulk properties at {$x$} $\sim$ 0.0 and {$x$} $\sim$ 1.0 essentially arise from the CB and VB electrons, respectively, and (ii) the bulk can be insulating at 0.3 $\lesssim x \lesssim$ 0.65. It is noted here that the bulk-insulating $x$ region determined from the resistivity measurements is 0.5 $\lesssim x \lesssim$ 0.9, which is shifted as a whole toward higher $x$ region by $\sim$0.2. This is likely due to the upward surface band bending with the order of $\sim$0.1 eV, as one can infer by referring to the location of chemical potential in Fig. 4(a).

Another important issue is the relationship between the location of the Dirac-cone SS and the bulk bands. As shown in Fig. 4(a), the energy position of the DP is lower than the VB top by $\sim$0.1 eV at $x$ = 0.0, as also illustrated in the band diagram in Fig. 4(b). Such an embedded nature of the DP persists up to $x$ = 0.4. On the other hand, the DP is isolated from the VB at $x$ = 0.6$-$1.0, satisfying a requirement for achieving the surface transport with the chemical potential located near the DP. Also, by taking into account that the bulk chemical potential seen in the ARPES data is shifted down by $\sim$0.1 eV compared to the true bulk chemical potential, it is inferred that the surface transport near the DP with insulating bulk can be achievable at $x$ {$\sim$} 0.6$-$0.8 in TBST [see Fig. 4(a)].

 Finally, we comment on the implications of the present observation in relation to the physical properties and device applications. As demonstrated in Fig. 2(c), the Dirac velocity of TBST is the largest among known TI solid solutions. This would be useful for realizing higher-performance electric devices which utilize the quantum transport of Dirac carriers. In addition, as mentioned above, the surface chemical potential of TBST was found to be robust against surface aging. This situation is different from BSTS \cite{ArakaneNC} and prototypical TIs \cite{FrantzeskakisPRB2015}, where the surface chemical potential is sensitive to small perturbations such as the gas adsorption on the surface, pointing to a good surface controllability in TBST, which is a prerequisite for realistic applications.
  
    In conclusion, we reported a comprehensive ARPES study of TBST. We have experimentally determined the evolution of bulk and surface electronic states upon Sb substitution and found (i) the sign reversal of surface Dirac carriers, (ii) the isolated Dirac-cone signature while keeping an insulating bulk at $x$ {$\sim$} 0.6, (iii) the largest Dirac velocity of the SS among TI solid solutions, and (iv) a high tunability of the chemical potential up to $\sim$0.6 eV. These observations strongly suggest that TBST is a promising material for the Dirac-cone engineering and device applications of TIs.

\begin{acknowledgements}
We thank Y. Tanaka, H. Kimizuka, and D. Takane for their assistance in the ARPES measurements. This work was supported by MEXT of Japan (Innovative Area ``Topological Materials Science'', Grant No. 15H05853), JSPS (KAKENHI Grants No. 15H02105, No. 26287071, No. 25287079, No. 25220708), KEK-PF (Grant No. 2015S2-003), and UVSOR (Grant No. 27-807).  

\end{acknowledgements}

\bibliographystyle{prsty}

\clearpage

{

\onecolumngrid
\begin{center}
{\large Supplemental Materials for \\
``Metal-insulator transition and tunable Dirac-cone surface state in topological insulator TlBi$_{1-x}$Sb$_x$Te$_2$ studied by ARPES'}

\vspace{0.3 cm}

Chi Xuan Trang,$^1$ Zhiwei Wang,$^{2,3}$ Keiko Yamada,$^1$ Seigo Souma,$^4$ Takafumi Sato,$^1$ Takashi Takahashi,$^{1,4}$ Kouji Segawa,$^5$ and Yoichi Ando$^{2,3}$

{\footnotesize
$^1${\it Department of Physics, Tohoku University, Sendai 980-8578, Japan}
\newline
$^2${\it Institute of Scientific and Industrial Research, Osaka University, Ibaraki, Osaka 567-0047, Japan}
\newline
$^3${\it Institute of Physics II, University of Cologne, K$\ddot{o}$ln 50937, Germany}
\newline
$^4${\it WPI Research Center, Advanced Institute for Materials Research, Tohoku University, Sendai 980-8577, Japan}
\newline
$^5${\it Department of Physics, Kyoto Sangyo University, Kyoto 603-8555, Japan}
}

\end{center}

\twocolumngrid

\raggedbottom

\renewcommand{\thefigure}{S\arabic{figure}}
\setcounter{figure}{0}

\subsection{S1. Physical mechanism to control the chemical potential}
We comment on the physical mechanism to control the chemical potential. In TBST, because of the similar ionic radius of Bi$^{3+}$ and Te$^{2-}$ ions, the antisite defects for Bi/Te would occur relatively easily, as is already known for the Bi$_2$Te$_3$ system. Bi ions sitting on the Te sites act as acceptors, while Te ions sitting on the Bi sites are donors. Increasing the Sb composition would suppress these antisite defects, and such a suppression of the anitisite defects is most likely the essential mechanism to control the chemical potential in the present system. We suggest that the chemical potential of TBST can be also controlled for a given $x$, based on the experimental fact that the dominant type of the antisite defects switches depending on the growth condition ({\it i.e.} whether the composition of the melt is Bi/Sb rich or Te rich).

\subsection{S2. Determination of the Dirac-cone band dispersion}

\begin{figure}[b]
 \includegraphics[width=3 in]{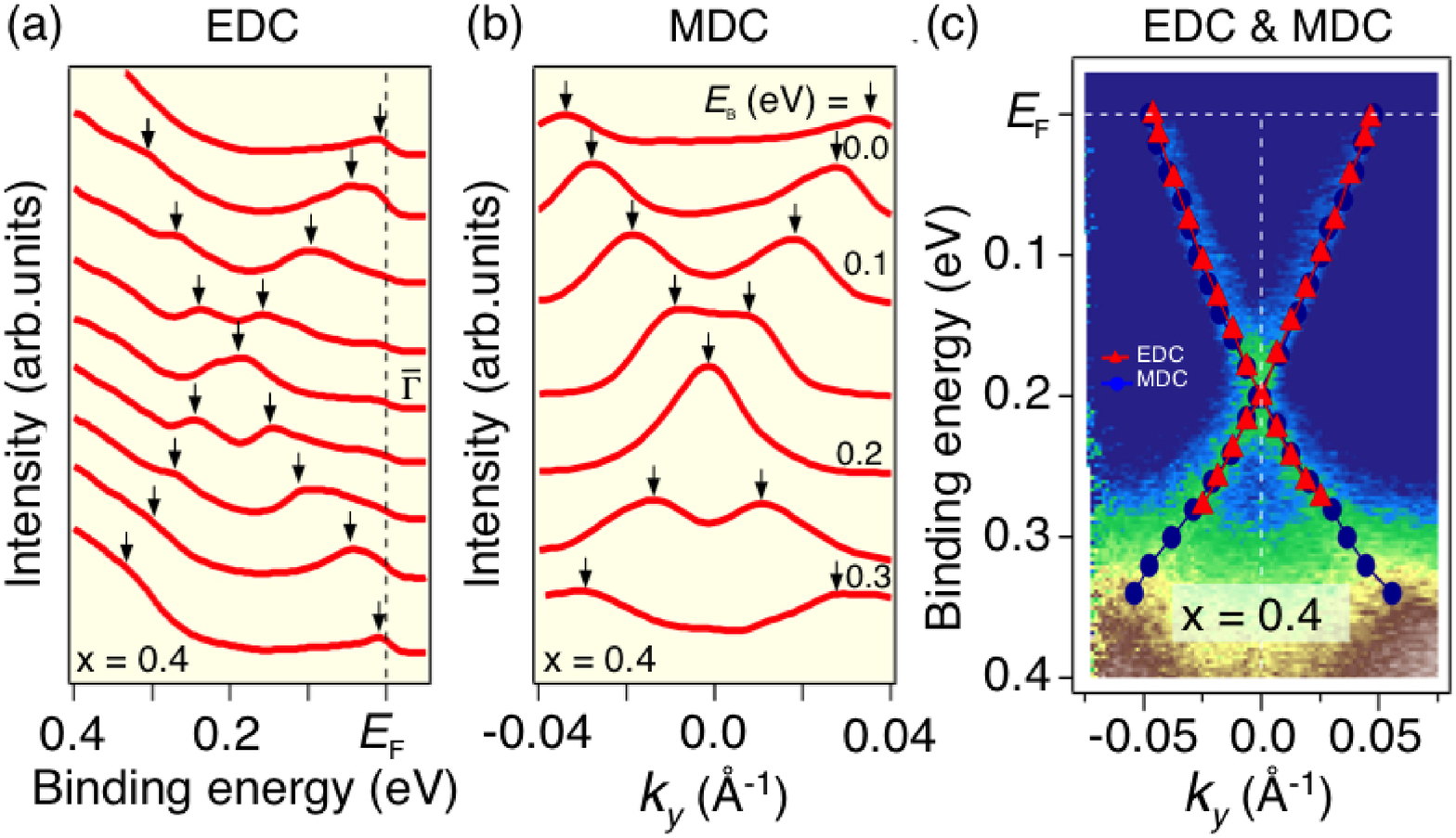}
\caption{(a), (b) Energy distribution curves (EDCs) and momentum distribution curves (MDCs) for $x$ = 0.4, respectively, measured with the Xe-I photons at $T$ = 30 K. Arrows indicate the band dispersion estimated from the peak position of the EDCs / MDCs. (c) ARPES intensity as a function of wave vector ($k_y$) and binding energy for $x$ = 0.4. Peak positions estimated from the EDCs and MDCs are overlaid with red triangles and blue circles, respectively.
}
\end{figure}

Figure S1(a) shows the energy distribution curves (EDCs) in the vicinity of the Fermi level ($E_{\rm F}$) around the $\bar{\Gamma}$ point for TBST ($x$ = 0.4) measured with the Xe-I photons ($h\nu$ = 8.437 eV). One can recognize two peaks away from the $\bar{\Gamma}$ point which merge into a single peak at the $\bar{\Gamma}$ point. This behavior is characteristic of a massless Dirac-cone surface state showing X-shaped band dispersion, as also visualized in the momentum distribution curves (MDCs) in Fig. S1(b). We found that the surface band dispersion estimated from the peak position of the EDCs in (a) well coincides with that estimated from the MDCs in (b), as displayed in Fig. S1(c). Since the EDC and the MDC analyses reflect essentially the same information on the band dispersion in the present study, we mainly used the EDCs for extracting the $x$-dependence of the Dirac-cone band dispersion in the main text (see Fig. 2).

\subsection{S3. Estimation of the conduction-band edge}

Since the bulk conduction band (CB) is well resolved with the He-I$\alpha$ line (compare Fig. S2(a) and Fig. 2(a) of the main text), we used the He-I$\alpha$ data to estimate the energy position of the CB bottom. According to the band calculation (ref. 23), the CB bottom is located at the $\Gamma$ point of the bulk Brillouin zone (BZ). In this regard, the He-I$\alpha$ line may not be best suited for accurately estimating the CB bottom since it basically probes the electronic states around the Z point of the bulk BZ. Nevertheless, according to the ARPES measurements of TlBiSe$_2$, the CB bottom estimated with the He-I$\alpha$  line and the $k_z$-tuned synchrotron light (see {\it e.g.}, Fig. 3 of ref. 24) reasonably overlap with each other despite the difference in the measured $k_z$ values, likely due to large $k_z$ broadening (about a half or more than a half of the $\Gamma$-Z length) originating from the short photoelectron escape depth. In such a case, it is more appropriate to use leading edge of the EDC at the $\bar{\Gamma}$ point, rather than to directly track the CB peak itself. As highlighted in Fig. S2(b), we have estimated the CB bottom from an intersection of two lines representing the leading edge of the CB and the spectral background. A possible deviation of the estimated CB bottom from the actual value is incorporated into error bar for $x$ = 0.0 in Fig. 4 of the main text.

\begin{figure}[h]
 \includegraphics[width=2.8 in]{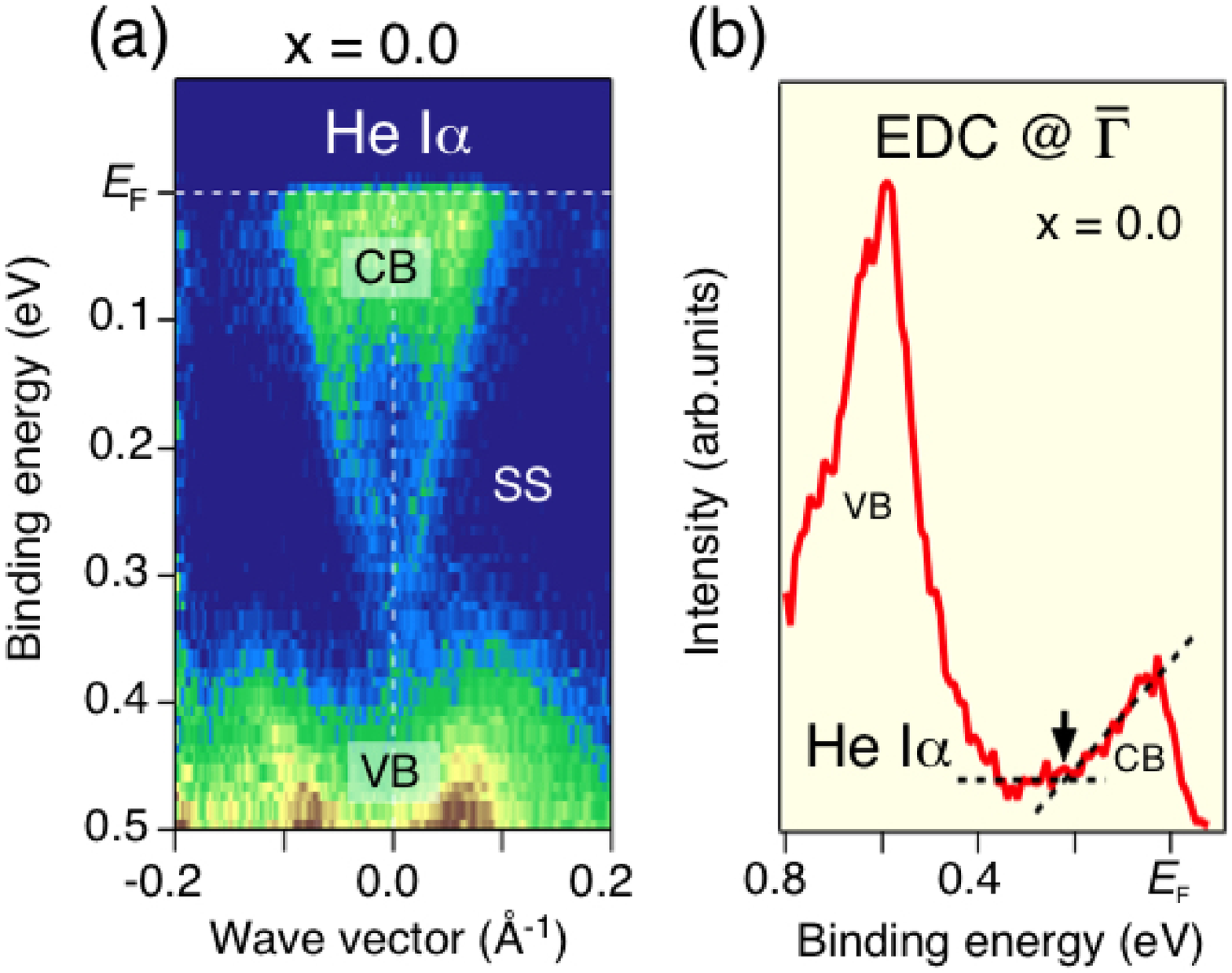}
\caption{(a) ARPES intensity along the $\bar{\Gamma}\bar{K}$ cut for $x$ = 0.0 measured with the He-I$\alpha$ photons at $T$ = 30 K. CB, SS, and VB denote the conduction band, surface state, and valence band, respectively. (b) EDC at the $\bar{\Gamma}$ point. Dashed lines on the EDC represent the leading edge of the CB and the spectral background. Their intersection corresponds to the location of the CB bottom (black arrow).
}
\end{figure}

\subsection{S4. Surface-aging experiment and cleaving plane}
We intended to control the surface chemical potential of TBST samples by depositing potassium (K) atoms onto the cleaved surface in ultrahigh vacuum. As visible in Fig. S3(a), the K deposition on the cleaved surface of Bi$_2$Se$_3$ bulk crystal leads to the apparent downward shift of the Dirac-cone band by $\sim$0.1 eV. On the other hand, the K deposition leads to no clear energy shift of the lower Dirac-cone band for the TBST surface ($x$ = 1.0). This demonstrates the robustness of surface chemical potential against intentional surface aging in TBST, in contrast to Bi$_2$Se$_3$.
   Here we comment on the cleaving plane of TBST. There exist two possible cleaving planes in TlBi$_{1-x}$Sb$_x$Te$_2$(TBST), {\it i.e.}, between Tl and Te layers, and between Te and Bi(Sb) layers (see Fig. 1(a) of the main text). Among these, TBST likely cleaves in the former way, judged from previous scanning tunneling microscopy and core-level photoemission spectroscopy studies on the cleaved surface of TlBiSe$_2$ (ref. 35). By also referring to the first-principles band-structure calculations (refs. 23 and 36), the ARPES-observed Dirac-cone state likely originates from the Te-terminated surface.

\begin{figure}[H]
\vspace{1cm}
\includegraphics[width=3in]{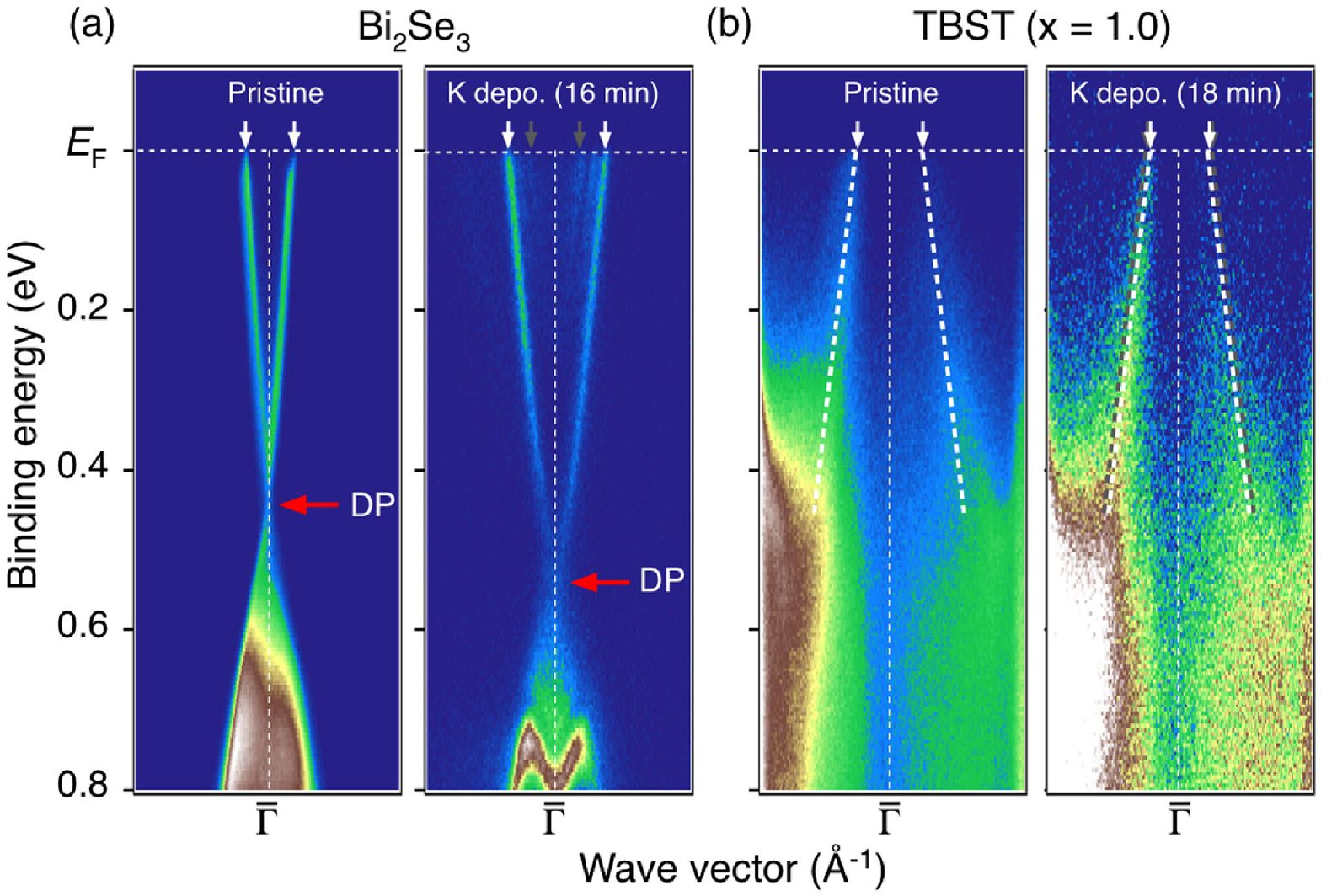}
\caption{(a) Comparison of the near-$E_{\rm F}$ ARPES intensity for fresh (left) and K-deposited surfaces of Bi$_2$Se$_3$ measured with $h\nu$ = 56 eV. (b) Same as (a) but for TBST ($x$ = 1.0). White arrows indicate the Fermi wave vector ($k_{\rm F}$), whereas red arrows denote the location of the DP. Gray arrows on right panel of (a) indicate the $k_{\rm F}$ point for the fresh surface of Bi$_2$Se$_3$. White dashed lines in (b) are guide to the eyes to trace the band dispersion. Gray dashed lines in right panel of (b) are the band dispersion for the fresh surface of TBST ($x$ = 1.0).
}
\end{figure}

\subsection {REFERENCES}
{
\noindent
[23] H. Lin, R. S. Markiewicz, L. A. Wray, L. Fu, M. Z. Hasan, and A. Bansil, Phys. Rev. Lett. \textbf{105}, 036404 (2010).\newline
[24] T. Sato, K. Segawa, H. Guo, K. Sugawara, S. Souma, T. Takahashi, and Y. Ando, Phys. Rev. Lett. \textbf{105}, 136802 (2010).\newline
[35] K. Kuroda, M. Ye, E. F. Schwier, M. Nurmamat, K. Shirai, M. Nakatake, S. Ueda, K. Miyamoto, T. Okuda, H. Namatame, M. Taniguchi, Y. Ueda, and A. Kimura, Phys. Rev. B \textbf{88}, 245308 (2013).\newline
[36] T. Shoman, A. Takayama, T. Sato, S. Souma, T. Takahashi, T. Oguchi, K. Segawa, and Y. Ando, Nature Commun. \textbf{6}, 6547 (2015). 
}

}

\end{document}